\documentclass[english,twocolumn]{svjour3}
\usepackage{units}
\usepackage{textcomp}
\usepackage{amsmath}
\usepackage{graphicx}

\makeatletter
\smartqed  % flush right qed marks, e.g. at end of proof

\makeatother

\usepackage{babel}

\begin{document}

\title{Optogalvanic Spectroscopy of Metastable States in Yb$^{\text{+}}$}

\author{M.J. Petrasiunas\and E.W. Streed\and T.J. Weinhold\and B.G. Norton\and D.~Kielpinski}

\institute{M.J. Petrasiunas\and E.W. Streed\and B.G. Norton\and D. Kielpinski \at Centre for Quantum Dynamics, Griffith University, Brisbane 4111, QLD, Australia\\
Tel.: +61-7-3735-3758\\
Fax: +61-7-3735-4426\\
\email{mattpetras@gmail.com}\\
\and T.J. Weinhold \at Centre for Quantum Computer Technology, Department of Physics, University of Queensland, Brisbane 4072, QLD, Australia}

\date{Received: date / Accepted: date}
\maketitle
\begin{abstract}
The metastable $\mathrm{^{2}F_{7/2}}$ and $\mathrm{\mathrm{^{2}D_{3/2}}}$ states of Yb$^{\text{+}}$ are of interest for applications in metrology and quantum information and also act as dark states in laser cooling. These metastable states are commonly repumped to the ground state via the 638.6 nm $\mathrm{^{2}F_{7/2}}$ -- $\mathrm{^{1}D[5/2]_{5/2}}$ and 935.2 nm $\mathrm{\mathrm{^{2}D_{3/2}}}$ -- $\mathrm{^{3}D[3/2]_{1/2}}$ transitions. We have performed optogalvanic spectroscopy of these transitions in Yb$^{\text{+}}$ ions generated in a discharge. We measure the pressure broadening coefficient for the 638.6 nm transition to be $\mathrm{70\text{\textpm}10\; MHz\cdot mbar^{-1}}$. We place an upper bound of 375 MHz/nucleon on the 638.6 nm isotope splitting and show that our observations are consistent with theory for the hyperfine splitting. Our measurements of the 935.2 nm transition extend those made by Sugiyama et al, showing well-resolved isotope and hyperfine splitting \cite{Sugiyama-1995}. We obtain high signal to noise, sufficient for laser stabilisation applications \cite{Streed-DAVLL-2008}. \keywords{ytterbium \and ion \and metastable \and optogalvanic \and spectroscopy} \PACS{31.30.Gs \and 42.62.Fi \and 52.80.Hc \and 52.80.Yr} 
\end{abstract}

\section{Introduction\label{sec:Introduction}}

The singly ionised ytterbium (Yb) atom has come under close interest for applications in a number of fields, including ion trap quantum computing and trapped ion optical frequency standards. Both of these areas provide motivation for investigation of the atomic structure of ytterbium ions. In this work, we study transitions from metastable states in Yb$^{\text{+}}$ that are important to both of these applications. Transitions in atoms with complex electronic structure, such as $\mathrm{Yb^{+}}$ are a challenge to understand solely through theoretical calculations, thus necessitating experimental research to form an accurate picture of these electronic transitions.

The $\mathrm{^{2}S_{1/2}-{}^{2}F_{7/2}}$ electric octupole transition in $\mathrm{Yb^{+}}$ is of particular interest for applications as an optical frequency standard \cite{Roberts-1997,Webster-2002,Blythe-2003,Hosaka-2005,Hosaka-2009}. The excited state spontaneous emission lifetime of the transition is estimated to be on the order of 6 years \cite{Hosaka-2005}. This remarkably long lifetime results in a natural transition linewidth on the order of nHz which provides an excellent basis for a highly accurate atomic clock transition. The clock laser excites ions into the long-lived $\mathrm{{}^{2}F_{7/2}}$ state, which is repumped to the ground state via the highly excited $\mathrm{^{1}D[5/2]_{5/2}}$ state. Taylor et al. investigated the alternative $3.43\unit{\mu m}$ $\mathrm{^{2}F_{7/2}}$ -- $\mathrm{^{2}D{}_{5/2}}$ transition, finding it to be many times weaker than the 638.6 nm $\mathrm{^{2}F_{7/2}}$ -- $\mathrm{^{1}D[5/2]_{5/2}}$ transition \cite{Taylor-1999}.%
\begin{figure}
\begin{centering}
\includegraphics[width=1\columnwidth]{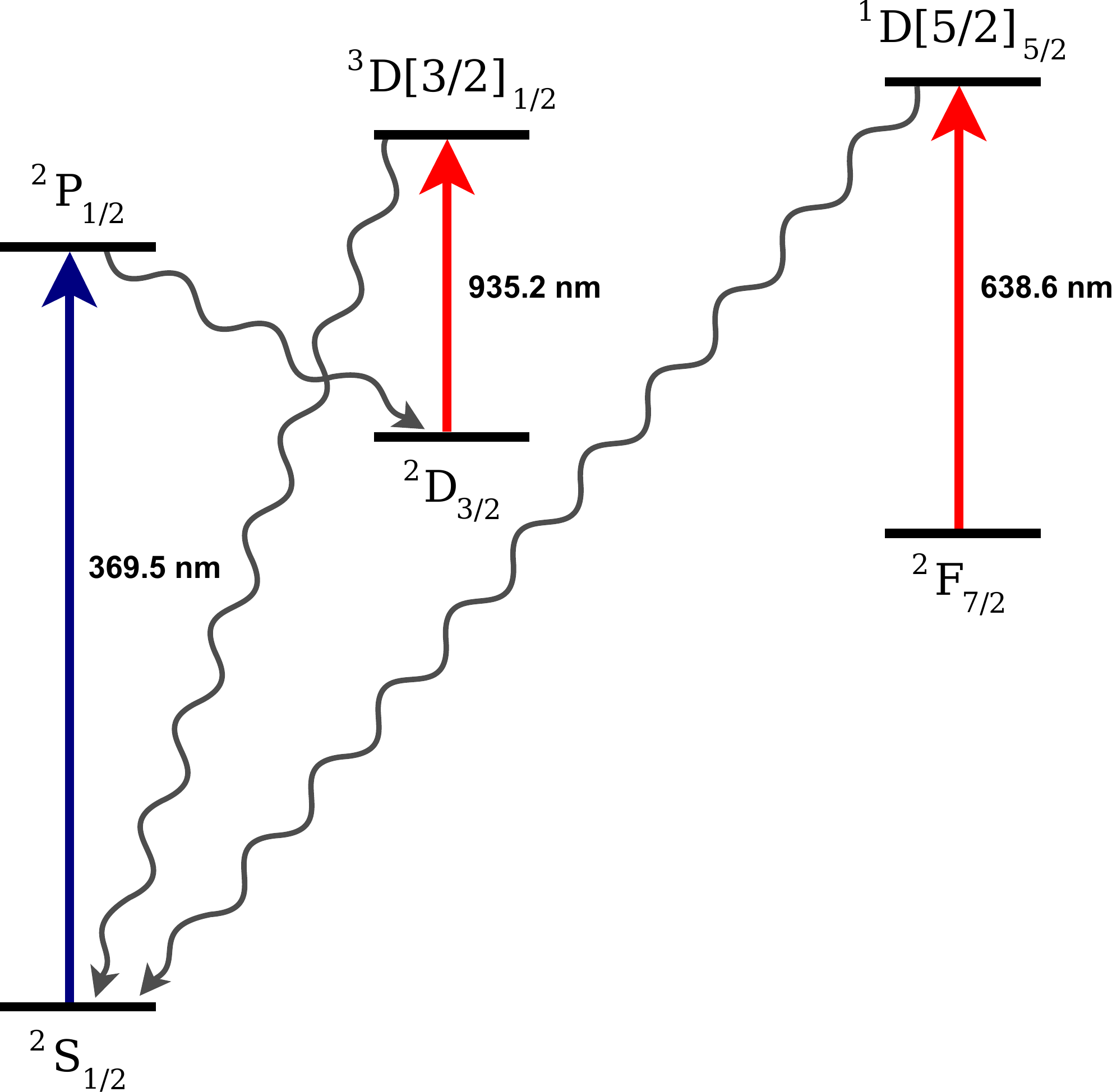}
\par\end{centering}

\begin{centering}
\caption{Energy level diagram for $\mathrm{Yb^{+}}$. The 638.6 nm $\mathrm{^{2}F_{7/2}}$ -- $\mathrm{^{1}D[5/2]_{5/2}}$ and 935.2 nm $\mathrm{\mathrm{^{2}D_{3/2}}}$ -- $\mathrm{^{3}D[3/2]_{1/2}}$ transitions studied in this work are shown in red. The 369.5 nm $\mathrm{^{2}S_{1/2}}$ -- $\mathrm{^{2}P_{1/2}}$ transition, used as principal laser cooling transition, has a 0.5\% branching ratio to decay to the metastable $\mathrm{^{2}D_{3/2}}$ state \cite{Yu-Maleki-2000}. During trap operation, collisions with residual background gas may also cause an ion to reach the metastable $\mathrm{^{2}F_{7/2}}$ state \cite{Olmschenck-2007}. Ions that fall into one of these {}`dark' states must be repumped to the ground state via the 638.6 nm or the 935.2 nm transition.}

\par\end{centering}

\centering{}\label{Flo:YbEnergyLevels}
\end{figure}

The principal laser cooling transition for $\mathrm{Yb^{+}}$ in an ion trap is the 369.5 nm $\mathrm{^{2}S_{1/2}}$ -- $\mathrm{^{2}P_{1/2}}$ transition. However the $\mathrm{^{2}P_{1/2}}$ state has a branching ratio of 0.5\% to decay into the metastable $\mathrm{^{2}D_{3/2}}$ state \cite{Yu-Maleki-2000}, necessitating repumping via the 935.2 nm $\mathrm{\mathrm{^{2}D_{3/2}}}$ -- $\mathrm{^{3}D[3/2]_{1/2}}$ transition to return the ion to the ground state, as shown in Figure \ref{Flo:YbEnergyLevels}. In general, trapped ions that are repumped by 369.5 nm and 935.2 nm light will be kept within the cooling cycle. However, over long periods of operation collisions with residual background gas may also cause an ion to reach the metastable $\mathrm{^{2}F_{7/2}}$ state \cite{Olmschenck-2007}. Ions in this state have fallen outside the range of the normal cooling cycle, and will remain in a {}`dark' state until returned to the $\mathrm{^{2}S_{1/2}}$ state \cite{Lehmitz-1989}. Despite the low rate of such occurrences, if the ions cannot be effectively isolated from collisions with an adequate vacuum pressure, then pumping of the $\mathrm{^{2}F_{7/2}}$ -- $\mathrm{^{1}D[5/2]_{5/2}}$ transition must be performed in order to return ions that reach this dark state to the cooling cycle. 

Due to the need for lasers to repump ions to the ground state from the metastable states discussed above, it is also important that these lasers can be stabilised to the transitions in question in order to counter the effects of frequency drift. We have previously demonstrated long-term absolute frequency stabilisation to ions in a discharge by employing the dichroic atomic vapor laser locking (DAVLL) method \cite{Streed-DAVLL-2008}.

The frequency shift of the transition due to the hyperfine structure and isotope of Yb$^{\text{+}}$ is important when tuning or stabilising a laser to the transition. Also, due to the difficulty of theoretical predictions of such shifts on excited to excited state transitions such as this one, experimental determination of the shifts can provide a useful test of the theoretical methods that can be used to predict them. Thus it is necessary to perform spectroscopy in order to determine the relevant isotope shift and hyperfine structure constants.

Optogalvanic spectroscopy is an alternative method for investigating ion transitions in a discharge. The electrical properties of the plasma are dependent on the electronic state populations of atoms and ions in the plasma. When the plasma is illuminated by a laser resonant with a transition, the change in plasma resistance yields a measurable signal. This method allows for electrical detection of atomic transitions, resulting in far greater sensitivity than optical methods such absorption. This provides a particular advantage in studying transitions from metastable states such as we do in this paper, where the spectroscopic signal can be quite weak due to the limited population of these states. The necessary metastable states can be populated by collisions in the discharge \cite{Hannaford-1983,Barbieri-1990}.

In this paper we measure the optogalvanic spectra of the 638.6 nm $\mathrm{^{2}F_{7/2}}$ -- $\mathrm{^{1}D[5/2]_{5/2}}$ and 935.2 nm $\mathrm{\mathrm{^{2}D_{3/2}}}$ -- $\mathrm{^{3}D[3/2]_{1/2}}$ transitions. We also provide a bound on isotope splitting for the 638.6 nm transition and show that the hyperfine constant is consistent with theory. We investigate the broadening processes present in both sets of spectra, eliminating pressure broadening of the 638.6 nm line in order to improve our measurement. In the case of the 935.2 nm transition, we obtain well-resolved isotopic and hyperfine lines, extending the measurements made by Sugiyama et al \cite{Sugiyama-1995}.

\section{Experimental Method\label{sec:Experimental Method}}

We performed optogalvanic spectroscopy in Yb hollow-cathode lamps containing Ne buffer gas at pressures of 6.7 mbar and 0.67 mbar. All hollow-cathode discharges used had the same geometry. The 6.7 mbar discharge was generated in a sealed Hamamatsu L2783-70NE-YB hollow cathode lamp, and was used to perform spectroscopy on both metastable transitions of interest.

A second purpose-built discharge was prepared inside a vacuum chamber, intended to allow for variation in the buffer gas pressure within the discharge with the purpose of operation at 0.67 mbar. This was done in order to see an expected $10\times$ reduction in pressure broadening. The chamber was evacuated directly by a rotary vacuum pump and vacuum pressure was monitored by a thermocouple gauge, sensitive down to $10^{-3}$ mbar. After evacuation, the chamber was baked at 180\textdegree{}C over a period of 2--3 days. At this point, a base pressure of 0.013 mbar was attained. A noble buffer gas was then able to be introduced via a needle valve. The buffer gas pressure was monitored with a thermocouple gauge.

\begin{figure*}
\begin{centering}
\includegraphics[width=1.6\columnwidth]{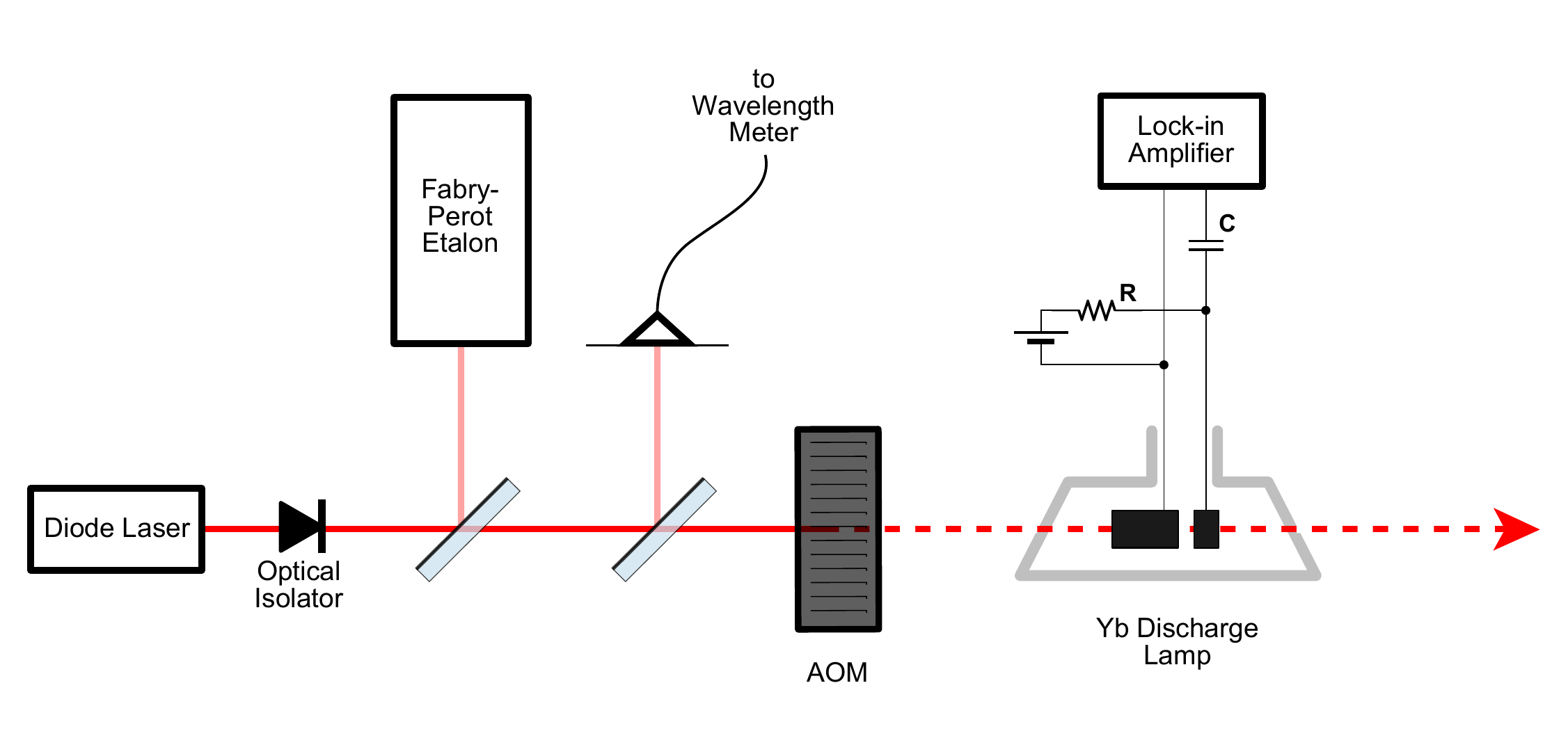}
\par\end{centering}

\begin{centering}
\caption{Schematic of setup used for optogalvanic spectroscopy. Laser diodes tuned by current and temperature to be on resonance with 638.6 nm and 935.2 nm transitions, with fine and course frequency reference from etalon and wavelength meter respectively. Optical isolation is present immediately after the laser. The laser beam is amplitude modulated by an acousto-optic modulator (AOM). Both 6.7 mbar and 0.67 mbar discharges were used. The signal due to laser excitation is filtered and demodulated by the lock-in amplifier.}

\par\end{centering}

\centering{}\label{Flo:LD:OGSSetup}
\end{figure*}

Sanyo DL5038-021 laser diodes were tuned for use at 638.6 nm using temperature and operating current. Behaviour close to the desired wavelength varied largely from diode to diode. The diodes were operated without an external cavity, running single-frequency with a linewidth less than 30 MHz, where the output was optically isolated to prevent optical feedback and damage to the diode. A Fabry-Perot etalon with an FSR of 300 MHz was used as a relative frequency reference to measure the frequency splittings and peak widths of the measured spectrum. Absolute wavelength measurement for tuning purposes was performed using a HighFinesse WS5-332 Wavemeter with $\pm0.001$ nm accuracy. The implementation of the setup for optogalvanic spectroscopy followed the schematic shown in Figure \ref{Flo:LD:OGSSetup}. %

The 935.2 nm spectroscopy setup was similar to that for the 638.6 nm line. A single mode tuneable external cavity laser was constructed with an Axcell Photonics laser diode (M9-935-0200-S30) and a 1200 g/mm holographic grating. The frequency was continuously tuneable over a range of 13.7 GHz by tuning the external cavity length with a piezo-electric actuator and synchronously varying the laser diode current. The laser spectral mode quality was monitored with a confocal Fabry-Perot spectrometer with a free spectral range of $478\pm8$ MHz and a finesse of 11. The observed spectrum bounds the 935.2 nm laser linewidth to below 50 MHz, though linewidths of a few MHz are typical for this laser configuration. The absolute wavelength was measured with a Burleigh WA-1100 Wavemeter (accuracy $\pm0.001$ nm). The 935.2 nm light was fibre coupled for delivery to the sealed discharge lamp. This provided 19 mW of 935.2 nm power at the discharge lamp in a laser beam spot with a diameter of $1.70\pm0.02$ mm (1/e$^{2}$), corresponding to a maximum intensity of $1.674\pm0.039$ $\unit{W}\cdot\unit{cm}^{-2}$.

In order for the optogalvanic signal to be distinguished from the electronic background noise, the optical beam was modulated at a set reference frequency using either amplitude or frequency modulation. We demodulated the signal using lock-in detection. A significant advantage of this method is that it allows for removal of $1/f$ noise, which would otherwise dominate the signal, allowing a high signal-to-noise ratio to be obtained. In these experiments we chopped the beam with an acousto-optic modulator (AOM), switching the beam intensity between zero and maximum intensity.

\section{Results for 638.6 nm $\mathrm{^{2}F_{7/2}}$ -- $\mathrm{^{1}D[5/2]_{5/2}}$ transition\label{sec:638 Results}}

\begin{figure*}
\begin{centering}
\includegraphics[width=1.6\columnwidth]{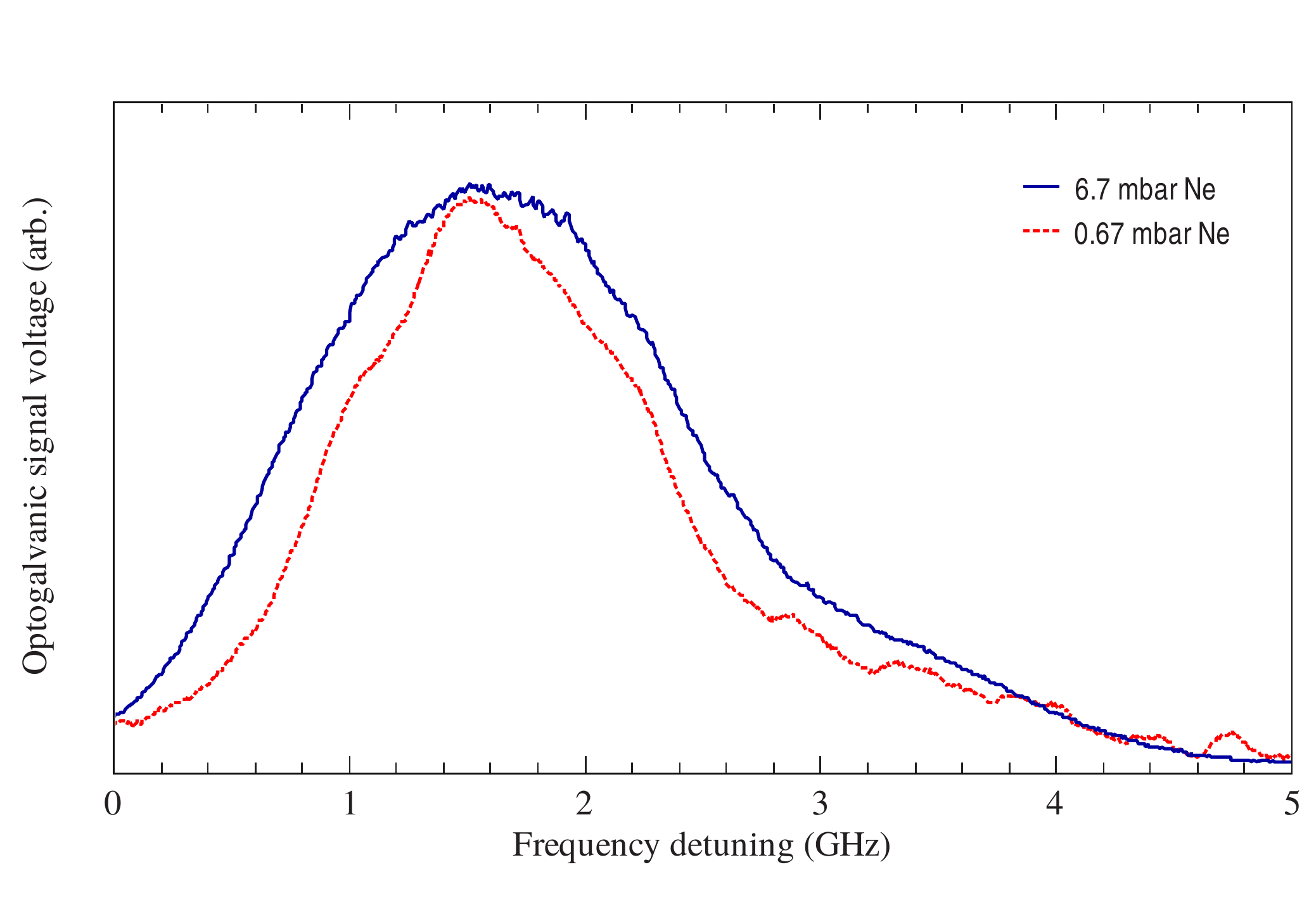}
\par\end{centering}

\centering{}\caption{\label{Flo:LD:OGSpectra1}Optogalvanic spectra of the 638.6 nm $\mathrm{^{2}F_{7/2}-^{1}D[5/2]_{5/2}}$ transition line in Yb$^{\text{+}}$, as observed in Yb$^{\text{+}}$ discharges with 6.7 mbar (blue line) and 0.67 mbar (red line) of Ne buffer gas. The spectra are normalised for ease of comparison. The reduction in signal linewidth from 1.9 GHz at 6.7 mbar to 1.5 GHz at 0.67 mbar is attributed to pressure broadening.}

\end{figure*}

We obtained optogalvanic spectra from discharge lamps operating with a discharge current of 3 mA at 6.7 mbar and 2 mA at 0.67 mbar, for which the maximum signal amplitude was obtained. The 638.6 nm laser power used at the lamp was 10--13 mW, chopped at a modulation frequency of 10 kHz. The signals for both the sealed 6.7 mbar discharge and the low-pressure 0.67 mbar discharge with Ne buffer gas are shown in Figure \ref{Flo:LD:OGSpectra1}.%

The linewidth of the signal obtained at a buffer gas pressure of 6.7 mbar was found to be approximately 1.9 GHz (FWHM). The observed linewidth decreases to approximately 1.5 GHz as the pressure is reduced to 0.67 mbar. There is some structure in the low-pressure peak that could be attributed to the splitting of the 638.6 nm line. The frequency axis of the plots is determined by taking the frequency spacing of Fabry-Perot transmission peaks to be equal to the Fabry-Perot FSR.%
\begin{figure*}
\begin{centering}
\includegraphics[width=1.6\columnwidth]{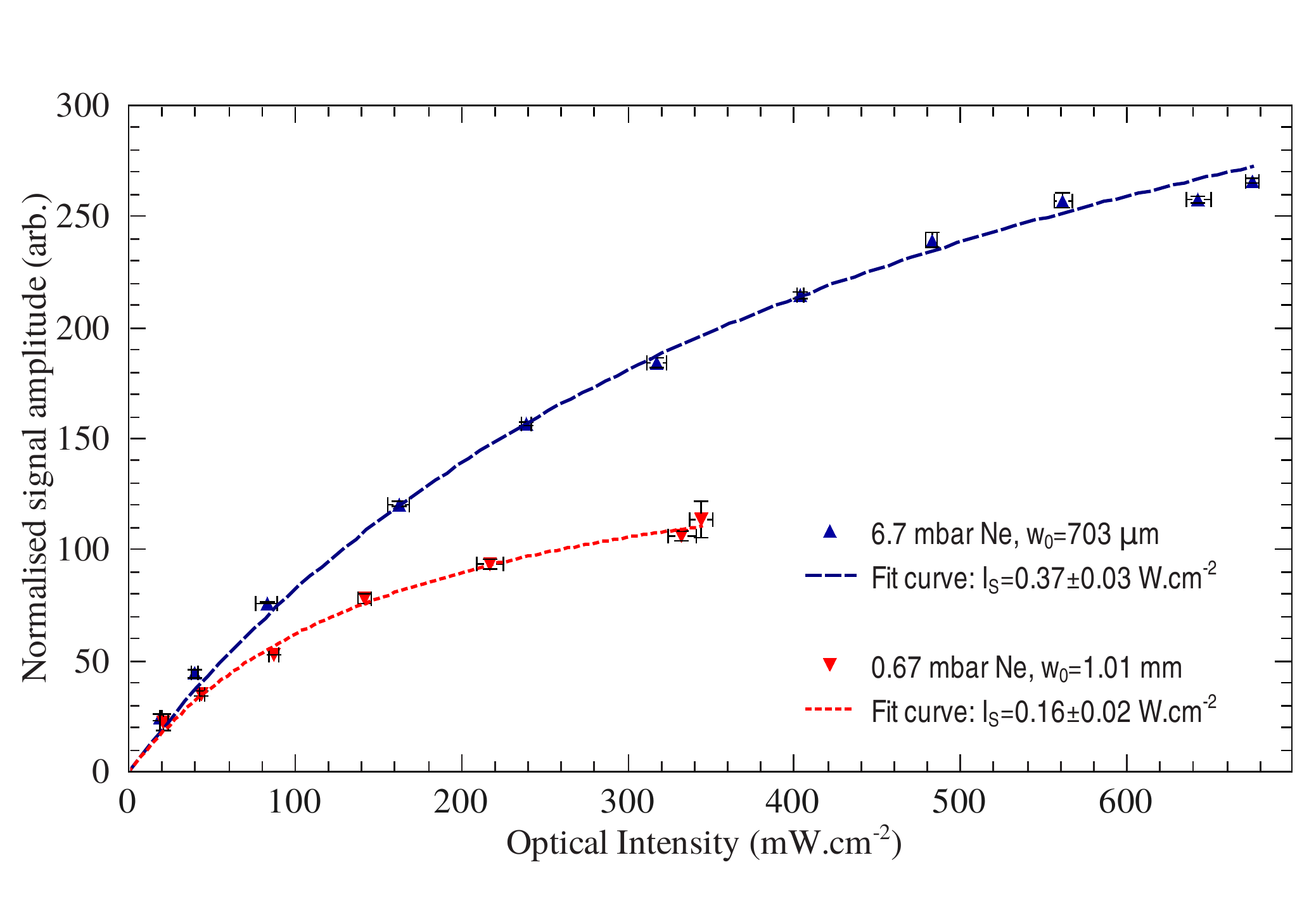}
\par\end{centering}

\centering{}\caption{\label{Flo:LD:SatCurves}The relationship of optogalvanic signal amplitude to optical intensity for both pressure levels. Blue colouring and upward triangles denote the measurements taken at 6.7 mbar Ne buffer gas pressure, whilst red and downward triangles denote the measurements taken at 0.67 mbar. The broken lines denote the fitted saturation curves. For ease of comparison, the signals are normalised to have equal slope near zero optical power.}

\end{figure*}

The optogalvanic signal amplitude was measured for a range of optical powers producing the saturation curve shown in Figure \ref{Flo:LD:SatCurves}. A pure pressure broadening model predicts a signal at maximum amplitude of the form \begin{equation}
V_{sig}\propto\frac{s}{1+s}\label{eq:SaturationEqn}\end{equation}
to describe the effect of saturation on the optogalvanic signal, where the saturation parameter $s=\frac{I}{I_{S}}$ is the ratio of optical intensity to the saturation value and $V_{sig}$ is the signal amplitude \cite{Metcalf-1999}. The saturation intensity $I_{S}$ is dependent on the homogeneous linewidth $\Gamma_{H}$ as shown by\begin{equation}
I_{S}=\frac{2\pi^{2}hc}{3\lambda^{3}}\Gamma_{H}\label{eq:LD:SatLinewidth}\end{equation}
where the homogeneous linewidth varies linearly with the pressure $P$ according to the Stern-Vollmer equation $\Gamma_{H}=\Gamma_{0}+\alpha P$ for constants $\Gamma_{0}$ and $\alpha$ \cite{Siegman-1986}.

Figure \ref{Flo:LD:SatCurves} shows experimental data on the saturation of optogalvanic response. The signals were fitted with Equation \eqref{eq:SaturationEqn} and the slopes near zero intensity have been normalised for ease of comparison. The measured optical power is combined with the beam waist size $w_{0}$ inside the discharge to give the optical intensity, and hence allow calculation of the saturation intensity $I_{S}$ for each measured curve.

A beam profiling camera was used to find the $1/e^{2}$ beam radius $w_{0}$ inside the 6.7 mbar discharge to be 730 $\mathrm{\mu m}$, resulting in an estimated saturation intensity of $0.37\pm0.03\;\mathrm{W\cdot cm^{-2}}$. The 0.67 mbar discharge had a saturation intensity of $0.16\pm0.02\;\mathrm{W\cdot cm^{-2}}$ from the fitted saturation curve, where the measured beam waist $w_{0}$ was 1.01 mm. As the effects of inhomogeneous broadening on the saturation behaviour cannot be ruled out, these values for the saturation intensity are upper bounds.

Using Equation \eqref{eq:LD:SatLinewidth} and the calculated saturation intensities gives maximum homogeneous linewidth estimates of $0.74\pm0.06\;\mathrm{GHz}$ for the 6.7 mbar discharge, and $\mathrm{0.32\pm0.04\; GHz}$ for the measurements of the 0.67 mbar discharge. The difference between the linewidths in the high- and low-pressure discharges is $\mathrm{0.42\pm0.08\; GHz}$, which agrees with the difference in total linewidths measured from the spectra. This gives values for the $\Gamma_{0}$ and $\alpha$ coefficients, such that $\Gamma_{0}=\mathrm{0.27\pm0.07\; GHz}$ and $\alpha=\mathrm{70\pm10\; MHz\cdot mbar^{-1}}$. The value for $\Gamma_{0}$ is the remaining broadening not induced by pressure, whilst $\alpha$ is a property of the discharge and buffer gas used.

The vacuum system used to construct the low-\linebreak{}
pressure discharge allowed for use of different buffer gases in place of Ne. Ar buffer gas was shown to generate Yb$^{+}$ ions, using a UV spectrometer to observe fluorescence from the 369.5 nm line to confirm this for our apparatus. However whilst Martensson-Pendrill et al. demonstrated optogalvanic spectroscopy of the 369.5 nm $\mathrm{^{2}S_{1/2}}$ -- $\mathrm{^{2}P_{1/2}}$ line using Ar buffer gas \cite{Martensson-Pendrill-1994}, we were unable to observe an optogalvanic signal from the 638.6 nm transition. No such issue occurred with a Ne buffer gas, allowing for both Yb$^{+}$ generation and the expected optogalvanic response. The strong Ne{*} transition line at 638.47 nm near the 638.6 nm Yb$^{+}$ transition did not appear to compromise the measurement. We observed a DC offset in the optogalvanic response at the 638.6 nm line, attributed to the wings of the extremely strong Ne{*} line.

\section{Results for 935.2 nm $\mathrm{\mathrm{^{2}D_{3/2}}}$ -- $\mathrm{^{3}D[3/2]_{1/2}}$ transition}

\begin{figure*}
\begin{centering}
\includegraphics[width=1.6\columnwidth]{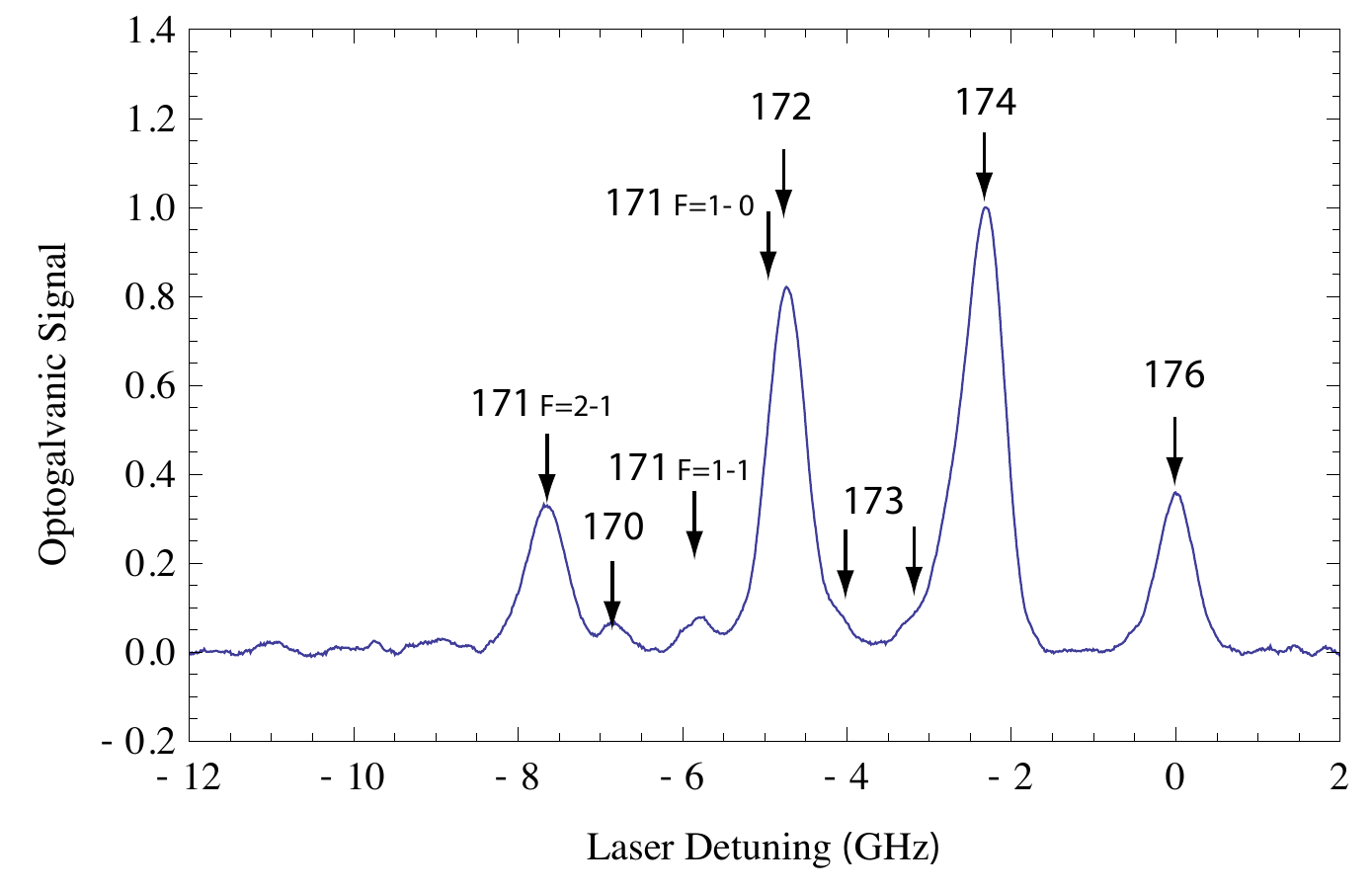}
\par\end{centering}

\centering{}\caption{\label{Flo:Spectrum935}Optogalvanic spectrum of the $\mathrm{\mathrm{^{2}D_{3/2}}}$ -- $\mathrm{^{3}D[3/2]_{1/2}}$ transition in Yb$^{+}$ at 935.2 nm in a sealed discharge lamp. Laser detuning is with respect to the Yb$^{\text{+}}$ 176 peak. The discharge lamp was run at its maximum rated current of 10 mA. The observed 176 Yb$^{\text{+}}$ peak FWHM linewidth is $535\pm7$ MHz. The laser illuminated the lamp with 19 mW of light and a maximum intensity of $1.674\pm0.039\;\mathrm{W\cdot cm^{-2}}$. Locations for Yb$^{\text{+}}$ 173 peaks are estimates from the total integrated signal weighting and the predicted location of the transition centroid from the isotopic shift trend.}

\end{figure*}

An optogalvanic spectrum with resolved isotope splittings was measured for the $\mathrm{\mathrm{^{2}D_{3/2}}}$ -- $\mathrm{^{3}D[3/2]_{1/2}}$ transition at 935.2 nm in the sealed discharge lamp. Overlapping spectra were taken to measure the response over the 25.4 GHz span from 935.159 to 925.233 nm. Figure \ref{Flo:Spectrum935} shows a 10 GHz span encompassing the observed peaks, taken with 18 mW of 935.2 nm laser power and amplitude modulated at 20 kHz. In these measurements, the optogalvanic signal increased monotonically with current, so the maximum rated lamp current of 10 mA was used. The four predominant peaks occurred at wavelengths of 935.173, 935.180, 935.187, and 935.197 nm. No peaks were observed outside of the region shown in figure \ref{Flo:Spectrum935} and the spectra is in qualitative agreement with previously published work \cite{Sugiyama-1995}. The hyperfine structure of Yb$^{\text{+}}$ 171 and 173 can be seen to overlap with the peaks from 174,172, and 170. From measurements in trapped ions \cite{McLoughlin-Hensinger-2011} the Yb$^{\text{+}}$ 172 transition is known to be only $-110$ MHz from the $F=1-F'=0$ transition of Yb$^{\text{+}}$ 171. Combined with the known hyperfine splitting of Yb$^{\text{+}}$ 171 allows the assignment of these features. A FWHM of $535\pm7$ MHz was measured on the 176 peak and is representative of the single isotope broadening. This peak is free from distortions induced by overlap between peaks due to the hyperfine splittings in Yb$^{\text{+}}$ 171 and 173. The integrated optogalvanic signal over the Yb$^{\text{+}}$ 176 peak is $12\pm1$\% of the total integrated signal, consistent with the natural abundance of 12.7\%. Other peak assignments were verified through a combination of known wavelengths and hyperfine splittings from trapped ion experiments and weighting based on natural abundance and Clebsch-Gordon coefficients. We measure a splitting between the 174/172 peaks of $2.39\pm0.01$ GHz and a splitting of $2.33\pm0.01$ GHz between the 174/176 peaks. This gives an average isotope shift of $1.18\pm0.03$ GHz/nucleon. The optogalvanic signal showed a linear dependence in laser power for 2.5 mW to 18 mW with no evidence of saturation. Changing the lamp current from 3 mA to 14 mA also produced a linear increase in signal. Similar optogalvanic spectra were observed with a blind hollow cathode lamp (Cathodeon 3UX Yb, Ne buffer gas). With a lock-in amplifier time constant of $\tau=300$ ms the maximum signal to noise observed was 58.%

\section{Discussion\label{sec:Discussion}}

Hyperfine and isotope splitting have strong effects on the optogalvanic lineshape. In the case of the 935.2 nm transition, the splitting due to these effects is well resolved. Previous spectroscopic work on this transition also gives a good idea of the composition of the expected spectrum \cite{Sugiyama-1995}.

In the case of the 638.6 nm transition, the broadening of the transition linewidth due to homogeneous and inhomogeneous sources makes resolution of the splitting difficult. Using theoretical estimates, the splitting spectrum that we expect to observe for the 638.6 nm in the well-resolved limit would contain 5 strong lines. This is based on the natural isotopic abundance of Yb$^{\text{+}}$ and the fact that the large number of hyperfine levels for $\mathrm{Yb^{+}}$ 173 ($I=\nicefrac{5}{2}$) will be far weaker in intensity than the transition lines for the 171, 172, 174 and 176 isotopes \cite{Hayden-1949}. We theoretically estimate the hyperfine splitting to be several GHz for $\mathrm{Yb^{+}}$ 171. The hyperfine splitting of the Yb$^{\text{+}}$ isotopes with even mass numbers will be zero, as these isotopes carry no nuclear spin.

Although broadening of the linewidth prevents clear resolution of hyperfine and isotope structure, deviation from the Gaussian shape on the higher-frequency side of the spectrum at the higher pressure of 6.7 mbar shows discernible structure in the observed peak. Hints of further structure become evident in the main peak at the lower pressure of 0.67 mbar, suggesting that this main peak is made up of several broadened lines.

As stated in Section \ref{sec:638 Results}, we obtained a $\mathrm{0.42\pm0.08\; GHz}$ difference between the maximum homogeneous linewidths in the 6.7 mbar and 0.67 mbar discharges from measurements of signal saturation. This agrees with the 0.4 GHz difference in total linewidth measured from the optogalvanic spectra, showing that the absolute difference is dominantly due to pressure broadening. However, as the achieved reduction in linewidth is significantly less than the expected factor of $10\times$ for pressure broadening, there is evidently another source of broadening. The residual pressure broadening at 0.67 mbar is $<50$ MHz, given that the expected linear relationship holds.

The contribution of power broadening on the line-width of the optogalvanic transition can be estimated to be $<100$ MHz, by estimating a natural linewidth on the order of 10 MHz. The lack of any significant contribution by power broadening was also confirmed by illuminating the discharge with powers of less than 1 mW and finding no noticeable difference in linewidth.

From the analysis above, it is evident that Doppler broadening is the only possible remaining broadening mechanism for the low-pressure measurements of the 638.6 nm line. The residual linewidth of the 638.6 nm transition appears to arise from the overlap of multiple unresolved Doppler-broadened lines. The temperature associated with a given Doppler linewidth is given by\begin{equation}
T=\frac{M\lambda^{2}\Gamma_{D}^{2}}{8k_{B}\ln2}\label{eq:DopplerTemp}\end{equation}
for atoms of mass $M$ and a Doppler-broadened FWHM linewidth $\Gamma_{D}$ \cite{Siegman-1986}. Results in the literature suggest that an expected temperature for a hollow-cathode discharge is 300 K, which would equate to a Doppler linewidth of approximately 440 MHz at 638.6 nm \cite{Hannaford-1983}. This level of Doppler broadening corresponds well with the optogalvanic spectrum in Figure \ref{Flo:LD:OGSpectra1} observed at 0.67 mbar, being at a level where multiple lines spaced by $<0.5$ GHz are not be fully resolved. This lends weight to the possibility of structure observed in the low-pressure optogalvanic spectrum.

Preliminary calculations of hyperfine structure were performed by Dr. W. M. Itano of the National Institute of Standards and Technology, Boulder, CO, USA. The hyperfine $A$ and $B$ constants of the $\mathrm{^{1}D[5/2]_{5/2}}$ state and the $\mathrm{^{2}F_{7/2}}$ state have been estimated by a multiconfiguration Dirac-Hartree-Fock method, utilising the GRASP package of atomic structure programs \cite{Parpia-1996,Jonsson-1996}. The method is similar to the cross-optimisation method used to calculate the hyperfine structure of the Hg$^{+}$ ion \cite{Brage-1999}, in which orbitals are optimised, stepwise, on different configurations. 

At each step, the previously optimised orbitals are held fixed. In the first step, the $\{1s,2s,2p,3s,3p,3d,4s,$\linebreak{}
$4p,4d,5s,5p\}$ core orbitals and $\{4f,6s\}$ valence orbitals are variationally optimised on the $J=(5/2,7/2)$ levels of the $4f^{13}6s^{2}$ configuration. Next, keeping the previously optimised orbitals fixed, the $\{5f,7s\}$ orbitals are determined by minimising the energy of the ground state, which is $4f^{14}6s$ in a first approximation. The $5d$ orbital is optimised on the $4f^{14}5d(J=3/2,5/2)$ levels. The $6p$ orbital is optimised on the $4f^{14}6p(J=1/2,3/2)$ levels. In the final step, the $\{8s,7p,6d,6f\}$ orbitals are optimised on the lowest-energy odd-parity $(J=9/2,11/2,13/2)$ levels.

This set of orbitals is used to generate a set of basis states by making single and double replacements from a set of reference configurations. The final approximate wavefunctions are calculated by relativistic configuration-interaction, i.e. by diagonalising the Dirac-Coulomb Hamiltonian matrix in this basis. The hyperfine constants are evaluated by evaluating the expectation values of the hyperfine energy operators in these wavefunctions. 

For the $\mathrm{^{1}D[5/2]_{5/2}}$ state, $A(\mathrm{Yb}\;171)=199$ MHz, $A(\mathrm{Yb}173)=-55$ MHz, and $B(\mathrm{Yb}\;173)=-1720$ MHz. For the $\mathrm{^{2}F_{7/2}}$ state, $A(\mathrm{Yb}\;171)=1105$ MHz, \linebreak{}
$A(\mathrm{Yb}\;173)=-304$ MHz, and $B(\mathrm{Yb\;}173)=-3680$ MHz. The only one of these constants for which there is an experimental measurement is $A(\mathrm{Yb}\;171)=905.0(0.5)$ MHz \cite{Taylor-1999}, measured for the $\mathrm{^{2}F_{7/2}}$ state, which is about 20\% lower than the calculated value. Our data is consistent with the splitting for the $\mathrm{^{1}D[5/2]_{5/2}}$ state being much smaller than the $\mathrm{^{2}F_{7/2}}$ state splitting, agreeing with Dr. Itano's theoretical estimate.

No theoretical results are available for the isotope shift of the 638.6 nm transition, but the normal mass shift is known to be small (on the order of a few MHz). The specific mass shift and field shift however are known to be far larger, providing the dominant contribution to the isotope shift. We can put an upper limit of approximately 375 MHz/nucleon on the isotope shift, given that there are at least three equally-spaced strong lines in the main peak with a combined linewidth of 1.5 GHz. The poorly resolved features in the low-pressure spectrum have approximately a 750 MHz spacing and could correspond to the isotopes 172, 174 and 176 of Yb$^{\text{+}}$.

The interpretation of the 935.2 nm transition optogalvanic spectra is substantially less complicated than that for 638.6 nm transition. The transition wavelengths for all isotopes excepting 168 (abundance 0.13\%) and 173 have previously been measured in trapped ion experiments. Our estimate of the isotope shift for the $\mathrm{\mathrm{^{2}D_{3/2}}}$ -- $\mathrm{^{3}D[3/2]_{1/2}}$ transition is in agreement with the $1.3\pm0.1$ GHz/nucleon from previous discharge lamp work \cite{Sugiyama-1995}. The small discrepancy between our measurement of $1.18\pm0.03$ GHz/nucleon and the $1.42\pm0.07$ GHz/nucleon from measurements in single trapped ions \cite{McLoughlin-Hensinger-2011} is most likely due to shifts induced in the lamp measurements by overlap of the 170, 172 and 174 isotope peaks with the hyperfine structure of Yb$^{\text{+}}$ 171 and 173. Agreement between the optogalvanic peaks and the published transition wavelengths for specific isotopes allows us to bound the overall lamp pressure shift as $<\pm60$ $\unit{MHz}\cdot\unit{mbar}^{-1}$, limited by the absolute accuracy of our wavemeter. We also place an upper limit of $80\pm1$ $\unit{MHz}\cdot\unit{mbar}^{-1}$ on the pressure broadening coefficient $\alpha$ for this transition, assuming that the linewidth of the Yb$^{\text{+}}$ 176 line is entirely due to pressure broadening. The prospects for long term frequency stabilisation to the 935.2 nm optogalvanic signal are encouraging given the observed signal to noise ratio of 58 for sub-second averaging.

\section{Conclusion\label{sec:Conclusion}}

We have measured the optogalvanic response of the 638.6 nm $\mathrm{^{2}F_{7/2}}$ -- $\mathrm{^{1}D[5/2]_{5/2}}$ and 935.2 nm $\mathrm{\mathrm{^{2}D_{3/2}}}$ --\linebreak{}
$\mathrm{^{3}D[3/2]_{1/2}}$ transitions in singly ionised Yb$^{\text{+}}$. We observe linewidths of 1.9 GHz and 1.5 GHz for the 638.6 nm optogalvanic peak, using discharges with 6.7 mbar and 0.67 mbar of Ne buffer gas, respectively, where Doppler broadening is expected to be the cause of these large linewidths and the inability to resolve the splittings. We performed measurements of the saturation curves in both pressure regimes, calculating the pressure broadening coefficient to be $\mathrm{70\text{\textpm}10\; MHz\cdot mbar^{-1}}$, and confirming that the source of the remaining broadening is inhomogeneous. The broadening prevents full resolution of transition lines in the main spectral peak, however our data is consistent with a much larger splitting of the $\mathrm{^{2}F_{7/2}}$ state than the $\mathrm{^{1}D[5/2]_{5/2}}$ state, as predicted by our theoretical estimate. We can provide an upper limit to the isotope splitting of approximately 375 MHz/nucleon. In our measurements of the 935.2 nm optogalvanic spectrum, we resolve with high signal-to-noise the peaks of the dominant isotopes of $\mathrm{Yb^{+}}$. We provide an upper bound of $80\pm1$ $\unit{MHz}\cdot\unit{mbar}^{-1}$ for the pressure broadening coefficient of this transition. The optogalvanic spectra observed in the 6.7 mbar discharge expand on the measurement made by Sugiyama et al \cite{Sugiyama-1995}. Our results demonstrate that a high signal to noise ratio can be achieved within a discharge for measurements of both transitions, also showing that both transitions are well suited for applications in laser stabilisation \cite{Streed-DAVLL-2008}.
\begin{acknowledgements}
We thank Dr. Wayne Itano for providing calculations of the 638.6 nm transition hyperfine structure. This work was supported by the U.S. Air Force Office of Scientific Research under AOARD Contracts FA2386-09-1-4132 and FA2386-10-1-4090 and by the Australian Research Council (ARC) under DP0773354 (Kielpinski), DP0877936 (Streed, Australian Postdoctoral Fellowship) and Howard Wiseman\textquoteright{}s Federation Fellowship FF0458313.\end{acknowledgements}


\begin{thebibliography}{21}
\bibitem{Sugiyama-1995}K. Sugiyama, J. Yoda, IEEE Trans. Instrum. Meas. \textbf{44}, 140 (1995)

\bibitem{Streed-DAVLL-2008}E.W. Streed, T.J. Weinhold, D. Kielpinski, Appl. Phys. Lett. \textbf{93}, 071103 (2008)

\bibitem{Roberts-1997}M. Roberts, P. Taylor, G.P. Barwood, P. Gill, H.A. Klein, W.R.C. Rowley, Phys. Rev. Lett. \textbf{78}, 1876 (1997)

\bibitem{Webster-2002}S.A. Webster, P. Taylor, M. Roberts, G.P. Barwood, P. Gill, Phys. Rev. A \textbf{65}, 052501 (2002)

\bibitem{Blythe-2003}P.J. Blythe, S.A. Webster, H.S. Margolis, S.N. Lea, G. Huang, S.-K. Choi, W.R.C. Rowley, P. Gill, R.S. Windeler, Phys. Rev. A \textbf{67}, 020501 (2003)

\bibitem{Hosaka-2005}K. Hosaka, S.A. Webster, P.J. Blythe, A. Stannard, D. Beaton, H.S. Margolis, S.N. Lea, P. Gill, IEEE Trans. Instrum. Meas. \textbf{54}, 759 (2005)

\bibitem{Hosaka-2009}K. Hosaka, S.A. Webster, A. Stannard, B.R. Walton, H.S. Margolis, P. Gill, Phys. Rev. A \textbf{79}, 033403 (2009)

\bibitem{Taylor-1999}P. Taylor, M. Roberts, G.M. Macfarlane, G.P. Barwood, W.R.C. Rowley, P. Gill, Phys. Rev. A \textbf{60}, 2829 (1999)

\bibitem{Yu-Maleki-2000}N. Yu, L. Maleki, Phys. Rev. A \textbf{61}, 022507 (2000)

\bibitem{Olmschenck-2007}S. Olmschenk, K.C. Younge, D.L. Moehring, D.N. Matsukevich, P. Maunz, C. Monroe, Phys. Rev. A \textbf{76}, 052314 (2007)

\bibitem{Lehmitz-1989}H. Lehmitz, J. Hattendorf-Ledwoch, R. Blatt, H. Harde, Phys. Rev. Lett. \textbf{62}, 2108 (1989)

\bibitem{Hannaford-1983}P. Hannaford, Contemp. Phys. \textbf{24}, 251 (1983)

\bibitem{Barbieri-1990}B. Barbieri, N. Beverini, A. Sasso, Rev. Mod. Phys. \textbf{62}, 603 (1990)

\bibitem{Metcalf-1999}H.J. Metcalf, P. Van der Straten, \textit{Laser Cooling and Trapping} (Springer Verlag, New York, 1999)

\bibitem{Siegman-1986}A.E. Siegman, \textit{Lasers} (University Science Books, Sausalito, California, 1986)

\bibitem{Martensson-Pendrill-1994}A. Martensson-Pendrill, D.S. Gough, P. Hannaford, Phys. Rev. A \textbf{49}, 3351 (1994)

\bibitem{McLoughlin-Hensinger-2011}J.J. McLoughlin, A.H. Nizamani, J.D. Siverns, R.C. Sterling, M.D. Hughes, B. Lekitsch, B. Stein, S. Weidt, W.K. Hensinger, Phys. Rev. A \textbf{83}, 013406 (1994)

\bibitem{Hayden-1949}R.J. Hayden, D.C. Hess, M.G. Inghram, Phys. Rev. \textbf{75}, 322 (1949)

\bibitem{Parpia-1996}F.A. Parpia, C. Froese Fischer, I.P. Grant, Comput. Phys. Commun. \textbf{94}, 249 (1996)

\bibitem{Jonsson-1996}P. J\"onsson, F.A. Parpia, C. Froese Fischer, Comput. Phys. Commun. \textbf{96}, 301 (1996)

\bibitem{Brage-1999}T. Brage, C. Proffitt, D.S. Leckrone, Astrophysical Journal \textbf{513}, 524 (1999)
\end{thebibliography}
\end{document}